\documentclass[runningheads,a4paper]{llncs}
\usepackage{multirow}
\usepackage[english]{babel}

\usepackage{amsmath}
\usepackage{graphicx}
\usepackage{multirow}
\usepackage{float}
\usepackage{caption}
\usepackage{subcaption}
\usepackage{hyperref}
\usepackage{booktabs}
\usepackage{hhline}
\usepackage{bbm}

\graphicspath{{imgs}}

\begin{document}

\title{Incorporating Task-Specific Structural Knowledge into CNNs for Brain Midline Shift Detection}
\titlerunning{CNNs with task-specific knowledge for midline shift detection}

\authorrunning{M. Pisov, M. Goncharov et al} 

\author{
    Maxim Pisov  \inst{1, 2} \and
    Mikhail Goncharov \inst{2, 3}  \and 
    Nadezhda Kurochkina \inst{1}  \and 
    Sergey Morozov \inst{4} \and
    Victor Gombolevskiy \inst{4} \and
    Valeria Chernina \inst{4} \and
    Anton Vladzymyrskyy \inst{4} \and
    Ksenia Zamyatina \inst{4} \and
    Anna Chesnokova \inst{4} \and
    Igor Pronin \inst{5} \and
    Michael Shifrin \inst{5} \and
    Mikhail Belyaev\inst{1, 2}
}

\institute{
    Skolkovo Institute of Science and Technology, Moscow, Russia
    \and 
    Kharkevich Institute for Information Transmission Problems, Moscow, Russia
    \and
    Moscow Institute of Physics and Technology, Moscow, Russia
    \and 
    Center for Diagnostics and Telemedicine, Moscow, Russia
    \and 
    Burdenko Neurosurgery Institute, Moscow, Russia
    \\
    \email{m.belyaev@skoltech.ru}
}

\maketitle

\begin{abstract}
Midline shift (MLS) is a well-established factor used for outcome prediction in traumatic brain injury, stroke and brain tumors.
The importance of automatic estimation of MLS was recently highlighted by ACR Data Science Institute.
In this paper we introduce a novel deep learning based approach for the problem of MLS detection, which exploits task-specific structural knowledge. We evaluate our method on a large dataset containing heterogeneous images with significant MLS and show that its mean error approaches the inter-expert variability. Finally, we show the robustness of our approach by validating it on an external dataset, acquired during routine clinical practice.

\end{abstract}

\section{Introduction}

\begin{figure}
    \begin{center}
      \includegraphics[width=.95\linewidth]{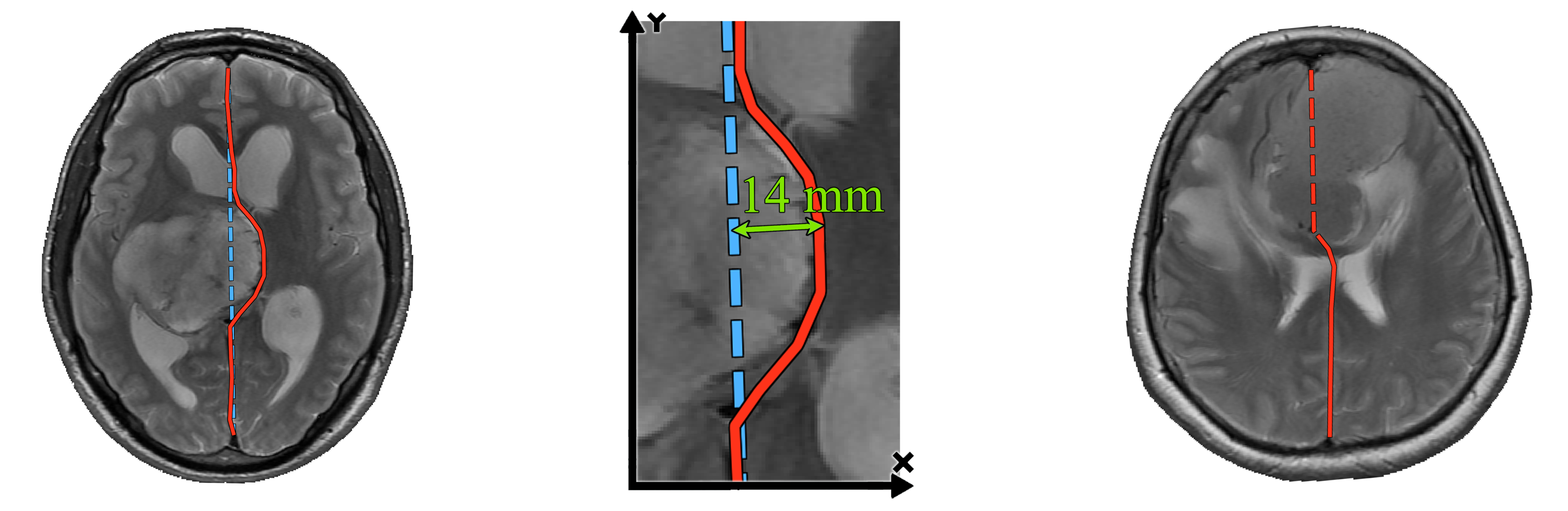}
      \caption{Left: an axial slice from a MRI image with corresponding midline (red) and a hypothetical normal midline (blue, dashed). Center: the midline shift. Right: a dubious case with an ill-defined midline (red, dashed).
      \vspace*{-6mm}
      }
      \label{fig:midline_sample}
    \end{center}
\end{figure}

The brain midline can be viewed as a line on axial and coronal projections of diverse imaging modalities (Fig. \ref{fig:midline_sample}, left). As the human brain is approximately symmetrical, the midline is straight in healthy subjects. However, various pathological conditions, such as traumatic brain injuries (TBI), stroke and brain tumors, may break this symmetry and lead to midline shift (MLS) \cite{liao2018brain}. 

A major number of studies show that MLS has a prognostic value for outcome prediction of various brain pathologies: level of consciousness in patients with acute intracranial hematoma \cite{ross1989brain}, 
median survival in patients with glioblastoma multiforme \cite{gamburg2000prognostic}, 
the outcome in patients with TBI \cite{jacobs2011computed}. Overall, early identification of patients with severe midline shift would assist patients management \cite{pullicino1997mass}. 
% In patients with TBI, MLS is associated with reduced cerebral metabolic rate of oxygen \cite{valadka2000midline}, as well as abnormal intracranial pressure \cite{eisenberg1990initial}.

However, definitions of significant MLS vary across studies. While the 5 millimeters (mm) threshold is frequently used, other approaches are common. For example, MLS larger than 9 mm was identified in \cite{pullicino1997mass}; the 5mm threshold was not justified within \cite{jacobs2011computed}. Such diversity is partly explained by the absence of a robust objective methodology of MLS estimation. A recent study \cite{paletta2018simplified} % of patients with middle cerebral artery stroke 
suggests that interrater variability of MLS estimation is rather high (intraclass correlation coefficients 0.72-0.89). 

The importance of MLS estimation and the need for its automation was recently highlighted by The American College of Radiology Data Science Institute \cite{mcginty2018acr}, and some promising results have already been achieved in this area (Section \ref{sec:previous}).
In this paper we propose a novel deep learning based approach\footnote{Full code for training and inference is available at GitHub: \\ \url{https://github.com/neuro-ml/midline-shift-detection}} for the MLS detection task.
We show that combining a standard segmentation approach with task-specific structural knowledge yields results which are more accurate, compared to straightforward CNNs for regression, and also interpretable, since the key part of the method is the midline localization. Moreover, we show that our method generalizes well on highly heterogeneous data and provide a natural way of estimating its confidence.

\section{Problem}
% \vspace*{-3mm}
\label{sec:problem}

We define the midline on an axial slice as a vertical curve that separates the brain hemispheres (Fig. \ref{fig:midline_sample}, left). 
The midline shift for an axial slice is then defined as the maximal distance between the midline (which might be deformed) and a hypothetical normal midline (Fig. \ref{fig:midline_sample}, center). Finally, the midline shift for a whole brain is the maximal midline shift across all axial slices where the midline is present.
\textbf{The task} is to determine, for a given brain image, the midline shift as well as the corresponding axial slice on which it is manifested.

It is worth noting that in some complicated cases 
even professional radiologists cannot confidently determine the localization of the midline (Fig. \ref{fig:midline_sample}, right).
Taking into account such dubious cases, it is also desirable that the method for MLS detection has a means of estimating its own confidence.

\section{Related work}
\label{sec:previous}

 Most of the methods for automatic MLS estimation are computer vision (CV) based and rely on keypoints detection. The proposed approaches often have a lot of "moving parts" which makes them hard to implement and fine-tune. For example, in \cite{midline_cv_falx} the authors use a four-step pipeline (edge detection, morphological filtering, lines detection, rule-based filtering) just to detect the cerebral falx. Another drawback of keypoints-based methods is that they require various important regions to be present on the image, e.g. many methods can be applied only to slices that contain ventricles \cite{midline_cv_ventricles} which makes them inapplicable to cases where the midline shift is manifested on lower or higher slices.

% Among other interesting approaches is the work \cite{midline_genetic} in which the authors regard the midline on a given axial slice as a quadratic Bézier curve and use a genetic algorithm to minimize a loss function across all midline points.

There are also a few papers that propose deep learning methods. In \cite{midline_resnet} the authors trained an adapted a version of ResNet to classify whether there is a significant midline shift on a given slice. Another interesting approach that combines deep learning with classical CV is described in \cite{midline_unet}. Here the authors use a U-Net \cite{unet} architecture for brain extraction, cisterns and acute intracranial lesions segmentation, while MLS detection is based on keypoints.

\section{Method}
\label{sec:proposed}
A straightforward deep learning approach is to directly predict the MLS via a convolutional neural network. Following the authors of \cite{midline_resnet}, we tested a ResNet-based \cite{resnet} network which predicted the MLS for each axial slice of given image. The final prediction was obtained as the maximal MLS only among the slices that contained an annotated midline. However, even in such a simplified design (the model did not need to filter out the slices for which the MLS was undefined),
this method yields poor results as we show in Section \ref{sec:results}.

Our intuition behind this is that the midline shift is a very high-level concept: the network needs to learn to detect several keypoints located very far from each other (Fig. \ref{fig:midline_sample}), as well as take into account their relative positions. The latter is a particularly difficult task for convolutional neural networks due to their invariance to translation. 

\begin{figure}[!b]
    \begin{center}
      \includegraphics[width=1.\linewidth]{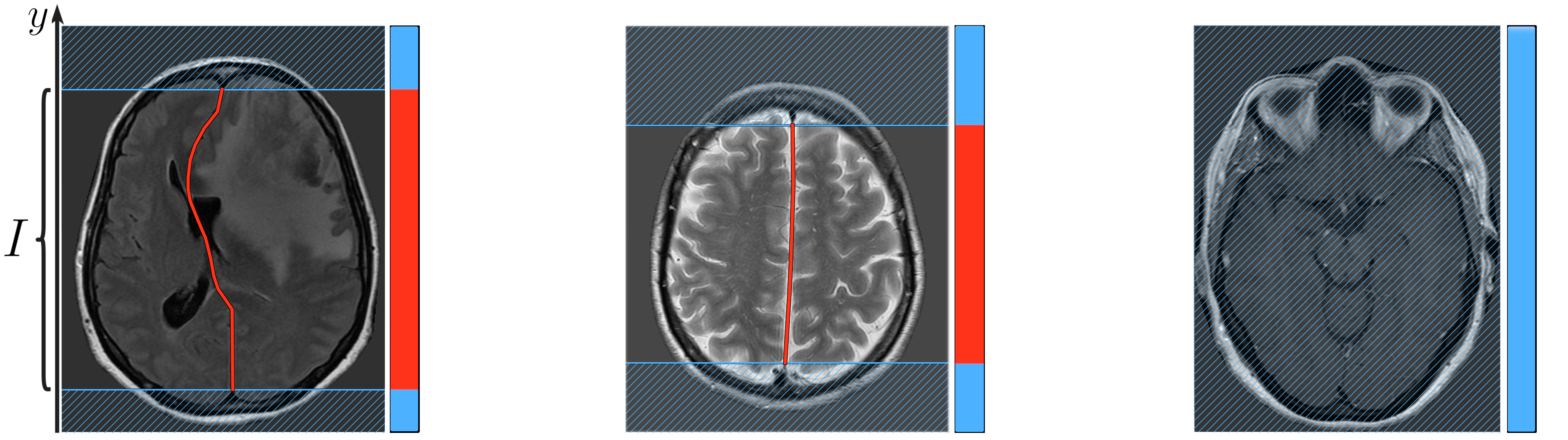}
      \caption{The binary masks of the regions where the midline is defined (red). Note the rightmost image, for which the midline is undefined everywhere.
      \vspace*{-5mm} 
      }
      \label{fig:limits}
    \end{center}
\end{figure}

On the contrary, the midline has visual features, like continuity and local symmetry, that are distinguishable on a smaller scale. This brings us to the idea to reduce the task of MLS prediction to the task of midline estimation: for a given slice we localize the midline while exploiting the structural knowledge about the target, then we derive the MLS from the predicted curve based on the definition given in Section \ref{sec:problem}. Normal midline is estimated as a straight line between prediction endpoints.

The key structural facts are: 1) for each coordinate $y$ there is at most one $x$-coordinate, which is refered as $\textbf{midline}_y$, such that the pixel $(\textbf{midline}_y, y)$ is situated on the midline; 2) $\textbf{midline}_y$ exists only for y-coordinates within certain interval $I$ on the $Oy$ axis to which binary mask we refer as \textbf{limits} (Fig.~\ref{fig:limits}).

\begin{figure}[b!]
    \begin{center}
      \includegraphics[width=1\linewidth]{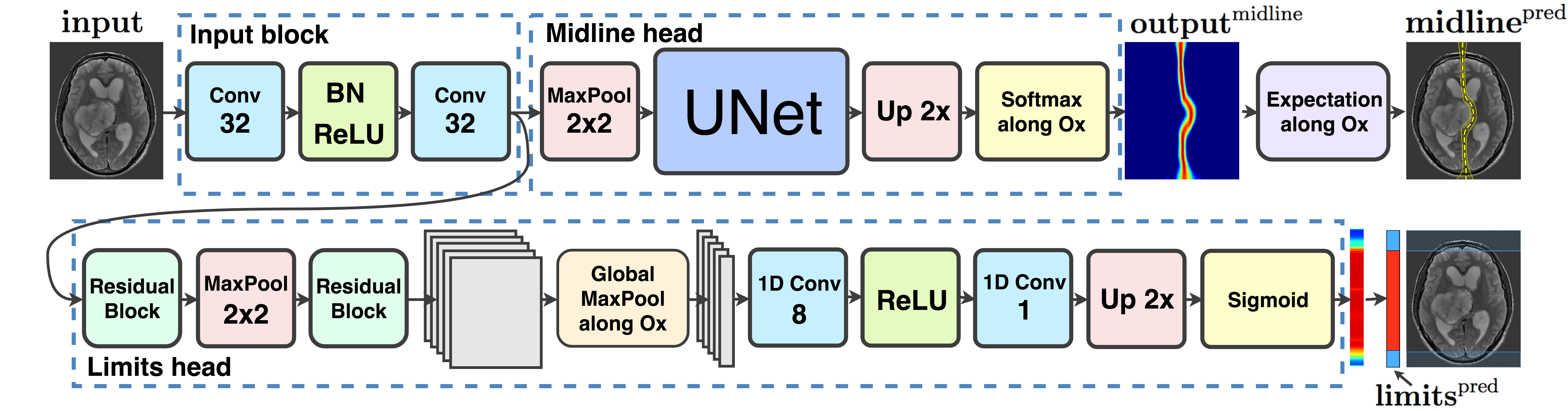}
      \caption{Schematic representation of the proposed architecture.
      \vspace*{-5mm}
      }
      \label{fig:architecture}
    \end{center}
\end{figure}

These facts imply that our method must be capable of solving the regression problem of \textbf{mildine} estimation and the classification problem of \textbf{limits} prediction.
To solve these tasks, we propose a two-headed convolutional neural network with shared input layers (Fig. \ref{fig:architecture}).
% and the detailed discussion of each head in the next subsections. 
As loss function, we optimize a weighted combination of standard losses for regression and classification:

\begin{multline*}
    \label{eq:final_loss}
    %L = \lambda_1 \cdot
    %\text{MSE}(\textbf{expectation}%[\textbf{limits}_{\text{true}}], \textbf{midline}[\textbf{limits}_{\text{true}}]) \\
    %+ \lambda_2 \cdot \text{BCE}(\textbf{limits}, \textbf{limits}_{\text{true}}),
    L = \lambda_1 \cdot \frac{1}{|I|} \sum_{y \in I} (\textbf{midline}_y - \textbf{midline}_y^{\text{pred}})^2 + \lambda_2 \cdot \text{BCE}(\textbf{limits}, \textbf{limits}^{\text{pred}}),
\end{multline*}
% \endgroup
where $\textbf{midline}_y^{\text{pred}}$ and  $\textbf{limits}^{\text{pred}}$ are the network's predictions, BCE is binary cross-entropy.

% \vspace*{-3mm}
\subsection{Midline estimation}
\label{sec:midline_estimation}

In order to 
% obtain midline coordinates 
estimate the midline
we adapt a segmentation approach. In a standard setting (with sigmoid activation and binary cross entropy loss) the output can be interpreted as "independent" probability of a particular pixel to be situated on the midline. In this case the \textbf{midline} is obtained after applying argmax along the $Ox$ axis. 

However, as we show in Section \ref{sec:results}, significantly better results can be achieved while imposing the following constraint on the output probability map 
\begin{equation}
    \label{eq:constraint}
    \sum_x \textbf{output}_{xy}^{\text{midline}} = 1,
    %\begin{cases}
    %    1,& \text{if } \textbf{limits}_y=1;\\
    %    0,              & \text{otherwise.}
    %\end{cases}
\end{equation}
which follows from the structural fact 1). Next, taking into account that for any given $y$-coordinate the head's output represents a probability distribution, we propose to predict the midline as its expected value:
\begin{equation*}
    \label{eq:expectation}
    \textbf{midline}_y^{\text{pred}} = \sum_x x \cdot \textbf{output}_{xy}^{\text{midline}}.
\end{equation*}

The overall architecture for midline estimation is shown in Fig. \ref{fig:architecture} (top). 
For our experiments we chose a UNet-based \cite{unet} architecture as a de facto standard for medical image segmentation. We replaced plain convolutional layers by residual blocks \cite{resnet} which are considered to improve the performance, as suggested by \cite{resunet}. %, and were an important component in our experiments that helped to avoid gradient explosion.
Also, during feature maps concatenation we use linear interpolation %instead of cropping in order 
to make the output's shape equal to the input's shape.
Finally, we apply a softmax nonlinearity to the network's output along the $Ox$ axis (instead of sigmoid), which ensures that the constraint from (\ref{eq:constraint}) is respected.
Note that because the head's output represents a probability distribution, at inference time we can calculate various statistics based on this distribution, e.g. percentiles, which are needed to estimate confidence intervals. This is a very important aspect of our approach which gives us a natural means of estimating the model's uncertainty. % (see Sections \ref{sec:results} and \ref{sec:discussion} for a more detailed discussion).

\vspace*{-2mm}
\subsection{Limits prediction} 
Since the proposed midline estimation approach yields $\textbf{midline}_y^{\text{pred}}$ for all $y$-coordinates, we need to filter out the predicted values for the regions where the midline is not defined (Fig. \ref{fig:limits}, hatched).  The corresponding limits are obtained by thresholding the second head's output ($\textbf{limits}^{\text{pred}}$) and taking the convex hull.

The architecture of the second head is shown in Fig. \ref{fig:architecture} (bottom). It has the same input layers as the midline estimation network, which are followed by two residual blocks \cite{resnet}.
Next, a global max pooling is applied along the $Ox$ axis in order to reduce the dimensionality of the 2D feature maps to 1D.
Finally we apply two 1D convolutions followed by the sigmoid activation function.

\section{Experimental setup}

At train time in all of our experiments we used
Adam optimizer \cite{adam} with default parameters ($\beta_1 = 0.9, \beta_2 = 0.999$) and a learning rate of $10^{-3}$, which showed the best results on the validation set. We used equal ($\lambda_1 = \lambda_2 = 1$) weights in the final loss as we didn't notice any loss imbalance at train time.

Also, we applied a simple preprocessing in order to reduce the data variability: resampling the axial slices to a $0.5 \times 0.5$ mm pixel spacing, background removal by Otsu thresholding \cite{otsu} and intensity normalization to zero mean and unit variance. Additionally, at train time we used random flips along the Ox axis as a cheap data augmentation technique.

The training was performed on batches of size 40 (which was simply determined by the amount of available GPU memory), until the validation scores reached a plateau, which happened at approx. 32000 batches. For this reason we used 32000 iterations for all our experiments.

\section{Data}
\label{sec:data}

In our experiments we used data from two sources. 

The first dataset (DS1) consists of 352 MRI series that come from a neurosurgery hospital and belong to patients with severe brain damage caused by tumors: 64\% of the images have a significant midline shift ($\geq$ 5mm), the mean MLS is 7.8$\pm$5.0mm. 
The dataset was labeled by an experienced neuroradiologist (exp1) and three specialists with limited background in neuroradiology (exp2-4). Their inter- and intra-expert variability is shown in Tab. \ref{tab:surface}. We split this dataset using 5-fold cross-validation. For each fold, we additionally leave 8 images out the training set to form a validation set.

The second dataset (DS2) comes from an out-patient clinic and represents a homogeneous sample of 203 MRI series acquired in routine clinical practice.
For this dataset only the MLS is available but not the midline itself; only 8\% of images have a large MLS ($\geq$ 5mm), the mean MLS is 2.9$\pm$1.5 mm.
We use this dataset only for final models' quality assessment in a prospective fashion.

The series from both sources contain only axial slices but have various voxel spacings, ranging from $0.2\times0.2\times1$mm to $1\times1\times5$mm, and modalities: T1 (25\%), T2 (68\%) and FLAIR (7\%). The images were collected using scanners from GE/Siemens and Toshiba/Siemens for DS1 and DS2 respectively.

\section{Results}
\label{sec:results}

\subsection{Midline shift detection}
We compare the proposed method with a direct MLS regression via ResNet \cite{resnet} on two tasks: 1) MLS prediction; 2) significant MLS ($\ge 5$mm) detection. 
In order to evaluate the quality of both methods we use mean absolute error (MAE) and the area under the ROC-curve (ROC AUC) for task 1 and 2 respectively. 
The ROC-curve was obtained by thresholding the predicted MLS by different values (from 0 to maximal MLS magnitude). The results are presented in Tab. \ref{tab:shift}.

\begin{table}
\vspace*{-6mm}
\centering
\caption{\label{tab:shift} Midline shift detection scores for various models ($\pm$ std) calculated on 5-fold cross-validation.}

\begin{tabular}{l|c|c|c|c}
                                 & \multicolumn{2}{c|}{MAE, mm} & \multicolumn{2}{c}{ROC AUC} \\ \cline{2-5} 
                                 & DS1         & DS2        & DS1           & DS2          \\ \hline
\multicolumn{1}{l|}{ResNet-152} & 2.92 $\pm$ 3.15 & 1.84 $\pm$ 1.10 & 0.91 $\pm$ 0.03 & 0.80 $\pm$ 0.04            \\ \hline
Proposed                         & \textbf{1.54 $\pm$ 1.98} & \textbf{0.75 $\pm$ 0.04} & \textbf{0.95 $\pm$ 0.02} & \textbf{0.92 $\pm$ 0.02}         
\end{tabular}
\vspace*{-6mm}
\end{table}

\subsection{Midline estimation}

\begin{table}
\centering
\caption{\label{tab:surface} Top: midline estimation metrics ($\pm$ std) calculated on 5-fold cross-validation for DS1. Bottom: neuroradiologist (exp1) variability on DS1.}
\begin{tabular}{l | c | c | c | c}
& MAX & RMSE & MAXs & RMSEs \\ 
\hline

Segmentation & 7.45 $\pm$ 9.84 & 0.95 $\pm$ 0.61 & 2.12 $\pm$ 3.25 & 0.81 $\pm$ 0.64 \\

Proposed & 3.61 $\pm$ 2.62 & 0.79 $\pm$ 0.44 & 1.58 $\pm$ 1.39 & 0.69 $\pm$ 0.54 \\
\hhline{=====}

exp1 vs exp1 & 3.16 $\pm$ 2.16 & 0.66 $\pm$ 0.19 & 1.47 $\pm$ 1.05 & 0.62 $\pm$ 0.50 \\
exp1 vs exp2-4 & 3.44 $\pm$ 2.13 & 0.77 $\pm$ 0.35 & 1.47 $\pm$ 0.97 & 0.66 $\pm$ 0.19 \\
\end{tabular}
\vspace*{-4mm}
\end{table}

In order to assess the midline estimation performance we use root-mean-square error (RMSE) as well as maximal error (MAX):
\begin{equation*}
\text{RMSE}(\textbf{midline}_y, \textbf{midline}_y^{\text{pred}}) = \sqrt{|I|^{-1}\sum\nolimits_{y \in I} (\textbf{midline}_y - \textbf{midline}_y^{\text{pred}})^2},
\end{equation*}
\begin{equation*}
\text{MAX}(\textbf{midline}_y, \textbf{midline}_y^{\text{pred}}) = \max\limits_{y \in I} |\textbf{midline}_y - \textbf{midline}_y^{\text{pred}}|.
\end{equation*}
These metrics, averaged along axial slices (MAXs, RMSEs) as well as entire brain images (MAX, RMSE), are shown in Tab. \ref{tab:surface}. 

We compare our method with a na\"ive segmentation approach mentioned in Section \ref{sec:midline_estimation}. Note that plain segmentation performs significantly worse in terms of maximal error, which is a more important characteristic for MLS detection.

\section{Discussion}
\label{sec:discussion}

\begin{figure}
    \vspace*{-7mm}
    \begin{center}
      \includegraphics[width=1\linewidth]{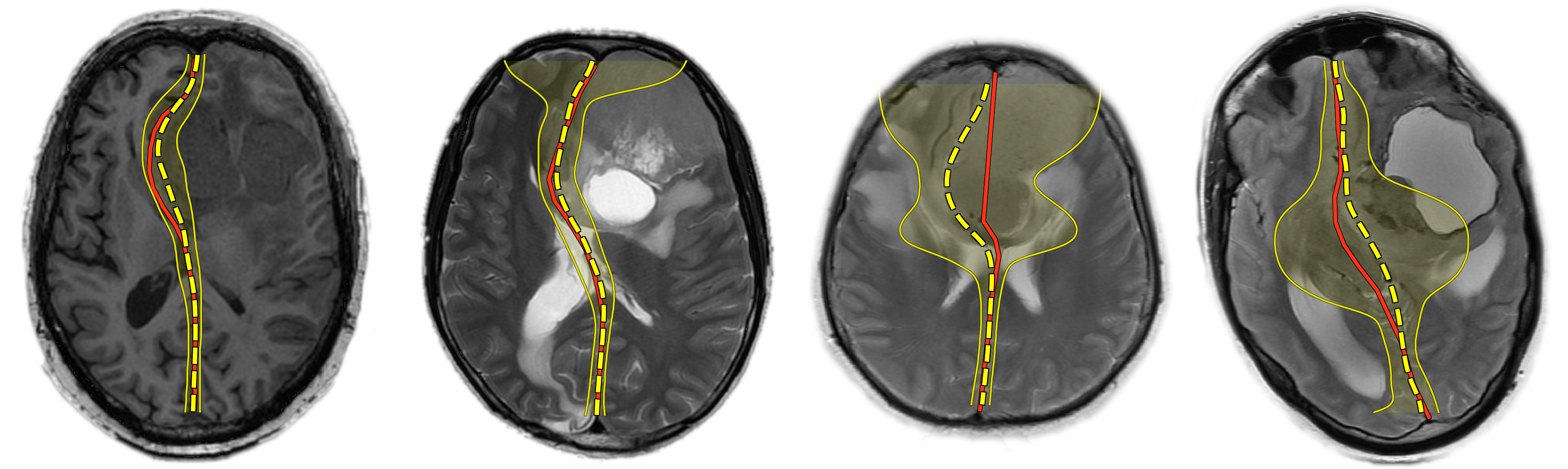}
      \caption{Ground-truth (red) and predicted (yellow, dashed) midlines with their 95\% confidence intervals for 2 random samples (left) and 2 typical examples from the set of cases with the largest errors (right).
      \vspace*{-7mm}
      }
      \label{fig:shifts}
    \end{center}
\end{figure}

Fig. \ref{fig:shifts} (right) shows several 
examples on which our method performs poorly. 
Our analysis of such examples suggests that the main source of errors are some really complicated cases that even professional radiologists have doubts with,  e.g.  images  on which  the  tumor  is  located  directly  in  the  middle  of  the brain, or incorrect cases with an extracerebral tumor located in the medial longitudinal fissure, e.g. falx meningioma. 
% However we have a natural way of detecting such cases - the width of the confidence intervals. 
Note how in the areas of greatest error the model's uncertainty is much higher.

Our preliminary experiments with CT images show that the proposed method can be easily adapted to work with CT, however we require a larger dataset to support this claim, which might be the subject of our future work.

\paragraph{Acknowledgements.} The development of the interpretable algorithm (done by M. Pisov and M. Goncharov) was supported by the Russian Science Foundation grant 17-11-01390.

%%% Moved by MB from Method section
%Note, that, because the first step yields the midlines for all slices, our method can be easily adapted to other approaches of measuring the midline shift \cite{liao2018brain}.

\bibliographystyle{splncs03}
\bibliography{main}

\begin{thebibliography}{10}
\providecommand{\url}[1]{\texttt{#1}}
\providecommand{\urlprefix}{URL }

\bibitem{midline_cv_ventricles}
Chen, W., Belle, A., Cockrell, C., Ward, K.R., Najarian, K.: Automated midline
  shift and intracranial pressure estimation based on brain ct images. Journal
  of visualized experiments: JoVE (74) (2013)

\bibitem{midline_resnet}
Chilamkurthy, S., Ghosh, R., Tanamala, S., Biviji, M., Campeau, N.G.,
  Venugopal, V.K., Mahajan, V., Rao, P., Warier, P.: Deep learning algorithms
  for detection of critical findings in head ct scans: a retrospective study.
  The Lancet  392(10162),  2388--2396 (2018)

\bibitem{gamburg2000prognostic}
Gamburg, E.S., Regine, W.F., Patchell, R.A., Strottmann, J.M., Mohiuddin, M.,
  Young, A.B.: The prognostic significance of midline shift at presentation on
  survival in patients with glioblastoma multiforme. International Journal of
  Radiation Oncology* Biology* Physics  48(5),  1359--1362 (2000)

\bibitem{resnet}
He, K., Zhang, X., Ren, S., Sun, J.: Deep residual learning for image
  recognition. In: Proceedings of the IEEE conference on computer vision and
  pattern recognition. pp. 770--778 (2016)

\bibitem{jacobs2011computed}
Jacobs, B., Beems, T., van~der Vliet, T.M., Diaz-Arrastia, R.R., Borm, G.F.,
  Vos, P.E.: Computed tomography and outcome in moderate and severe traumatic
  brain injury: hematoma volume and midline shift revisited. Journal of
  neurotrauma  28(2),  203--215 (2011)

\bibitem{midline_unet}
Jain, S., Vande~Vyvere, T., Terzopoulos, V., Maria~Sima, D., Roura, E., Maas,
  A., Wilms, G., Verheyden, J.: Automatic quantification of ct features in
  acute traumatic brain injury. Journal of Neurotrauma  (2019)

\bibitem{adam}
Kingma, D.P., Ba, J.: Adam: A method for stochastic optimization. arXiv
  preprint arXiv:1412.6980  (2014)

\bibitem{liao2018brain}
Liao, C.C., Chen, Y.F., Xiao, F.: Brain midline shift measurement and its
  automation: a review of techniques and algorithms. International journal of
  biomedical imaging  2018 (2018)

\bibitem{midline_cv_falx}
Liu, R., Li, S., Su, B., Tan, C.L., Leong, T.Y., Pang, B.C., Lim, C.T., Lee,
  C.K.: Automatic detection and quantification of brain midline shift using
  anatomical marker model. Computerized Medical Imaging and Graphics  38(1),
  1--14 (2014)

\bibitem{mcginty2018acr}
McGinty, G.B., Allen, B.: The acr data science institute and ai advisory group:
  harnessing the power of artificial intelligence to improve patient care.
  Journal of the American College of Radiology  15(3),  577--579 (2018)

\bibitem{resunet}
Milletari, F., Navab, N., Ahmadi, S.A.: V-net: Fully convolutional neural
  networks for volumetric medical image segmentation. In: 3D Vision (3DV), 2016
  Fourth International Conference on. pp. 565--571. IEEE (2016)

\bibitem{otsu}
Otsu, N.: A threshold selection method from gray-level histograms. IEEE
  transactions on systems, man, and cybernetics  9(1),  62--66 (1979)

\bibitem{paletta2018simplified}
Paletta, N., Maali, L., Zahran, A., Sethuraman, S., Figueroa, R., Nichols,
  F.T., Bruno, A.: A simplified quantitative method to measure brain shifts in
  patients with middle cerebral artery stroke. Journal of Neuroimaging  28(1),
  61--63 (2018)

\bibitem{pullicino1997mass}
Pullicino, P.M., Alexandrov, A., Shelton, J., Alexandrova, N., Smurawska, L.,
  Norris, J.: Mass effect and death from severe acute stroke. Neurology  49(4),
   1090--1095 (1997)

\bibitem{unet}
Ronneberger, O., Fischer, P., Brox, T.: U-net: Convolutional networks for
  biomedical image segmentation. In: MICCAI. pp. 234--241. Springer (2015)

\bibitem{ross1989brain}
Ross, D.A., Olsen, W.L., Ross, A.M., Andrews, B.T., Pitts, L.H.: Brain shift,
  level of consciousness, and restoration of consciousness in patients with
  acute intracranial hematoma. Journal of neurosurgery  71(4),  498--502 (1989)

\end{thebibliography}

\end{document}